\documentclass[aps,pra,twocolumn,groupedaddress]{revtex4-2}

\usepackage{amsmath}
\usepackage{amssymb}
\usepackage[pdftex]{graphicx}
\usepackage{multirow}
\usepackage{siunitx}
\usepackage[normalem]{ulem}

\begin{document}

\title{Fast manipulation of a quantum gas on an atom chip with a strong microwave field}

\author{Manon Ballu, Bastien Mirmand, Thomas Badr, H\'el\`ene Perrin and Aur\'elien Perrin}

\affiliation{Universit\'e Sorbonne Paris Nord, Laboratoire de Physique des Lasers,
CNRS UMR 7538, 99 av. J.-B. Cl\'ement, F-93430 Villetaneuse, France}

\date{\today}

\begin{abstract}
We report on an experimental platform based on an atom chip encompassing a coplanar waveguide which enables the manipulation of a quantum gas of sodium atoms with strong microwave fields. We describe the production with this setup of a very elongated degenerate quantum gas with typically $10^6$ atoms, that can be prepared all along the cross-over from the three-dimensional to the one-dimensional regime, depending on the atom number and trapping geometry. Using the microwave field radiated by the waveguide, we drive Rabi oscillations between the hyperfine ground states, with the atoms trapped at various distances from the waveguide. At the closest position explored, the field amplitude exceeds 5\,G, corresponding to a Rabi frequency on the strongest transition larger than \SI{6}{\mega\hertz}. This enables fast manipulation of the atomic internal state.
\end{abstract}

\maketitle

\section{Introduction}

Atom chips are a versatile technology for the manipulation of ultracold neutral atoms~\cite{Reichel1999,Haensel2001a,Reichel2002,Folman2002,fortagh:07,Reichel2011,Keil2016}. They primarily refer to surface-mounted structures enabling the production of magnetic traps of micrometric dimensions~\cite{Reichel2002,Folman2002,fortagh:07,Reichel2011,Keil2016}. Microfabrication gives access to a large variety of magnetic potentials resulting from the set of current-carrying wires. It also enables us to include waveguides radiating microwave fields near their surface \cite{Treutlein2004}. The latter has proven to be an effective tool for the coherent manipulation of the internal states of atoms with a hyperfine structure, such as alkali-metal atoms in their ground state \cite{Bohi2009,Bohi2010}, opening a wide range of applications such as compact atomic clocks \cite{Deutsch2010,Ramirez2011,Sarkany2014,Szmuk2015} or interferometers \cite{Bohi2009,Ammar2015,DupontNivet2015}, spectroscopy of the elementary excitations of a quantum gas \cite{Allard2016} or entanglement between quantum states \cite{Bohi2009,Colciaghi2023}. In these works, the microwave field is used either near resonance, driving coherent one-photon or two-photon Rabi oscillations with a long coherence time \cite{Treutlein2004,Deutsch2010}, or far from resonance to dress the atomic states \cite{Sarkany2014}.

Another promising application of microwave fields for neutral atoms on a chip concerns the manipulation of the collisional properties of a quantum gas. It has been proposed that a microwave field could be used to tune the scattering length of alkali-metal atoms near a microwave Feshbach resonance \cite{Papoular2010}. The field amplitude required for this application is, however, very large, on the order of several gauss. Microwave guides deposited on atom chips offer strong field amplitudes in the vicinity of the chip, providing an ideal platform for applications where a large field is required.

In this paper, we report on the manipulation of a sodium Bose-Einstein condensate confined in a magnetic microtrap with a strong microwave field produced by a coplanar waveguide in the immediate proximity of the atomic cloud. We observe fast coherent oscillations of the atomic states, driven by the large amplitude microwave field. We achieve a field amplitude exceeding 5\,G. These results pave the way to fast manipulation of atomic states for applications in quantum technologies.

The paper is organized as follows: In Sec.~\ref{at_ch_design}, we describe the atom chip design and present the different trapping potentials that can be obtained with our experimental platform. In Sec.~\ref{sec:bec}, we 
report on the production of a Bose-Einstein condensate with this setup and discuss its dimensionality, from three dimensions (3D) to one dimension (1D).
Finally in Sec.~\ref{fast_rabi}, we present the observation of fast coherent Rabi oscillations enabled by the strong microwave field. Additional technical details on the experimental sequence, 
the calibration of the static magnetic fields used to produce the magnetic trap, 
the calibration of atom number in absorption imaging, 
the transmission of the coplanar waveguide 
and 
the microwave coupling amplitude
are given in the appendixes.

\section{Experimental configuration}\label{at_ch_design}

\begin{figure}[t]
    \centering
    \includegraphics[width=0.85\linewidth]{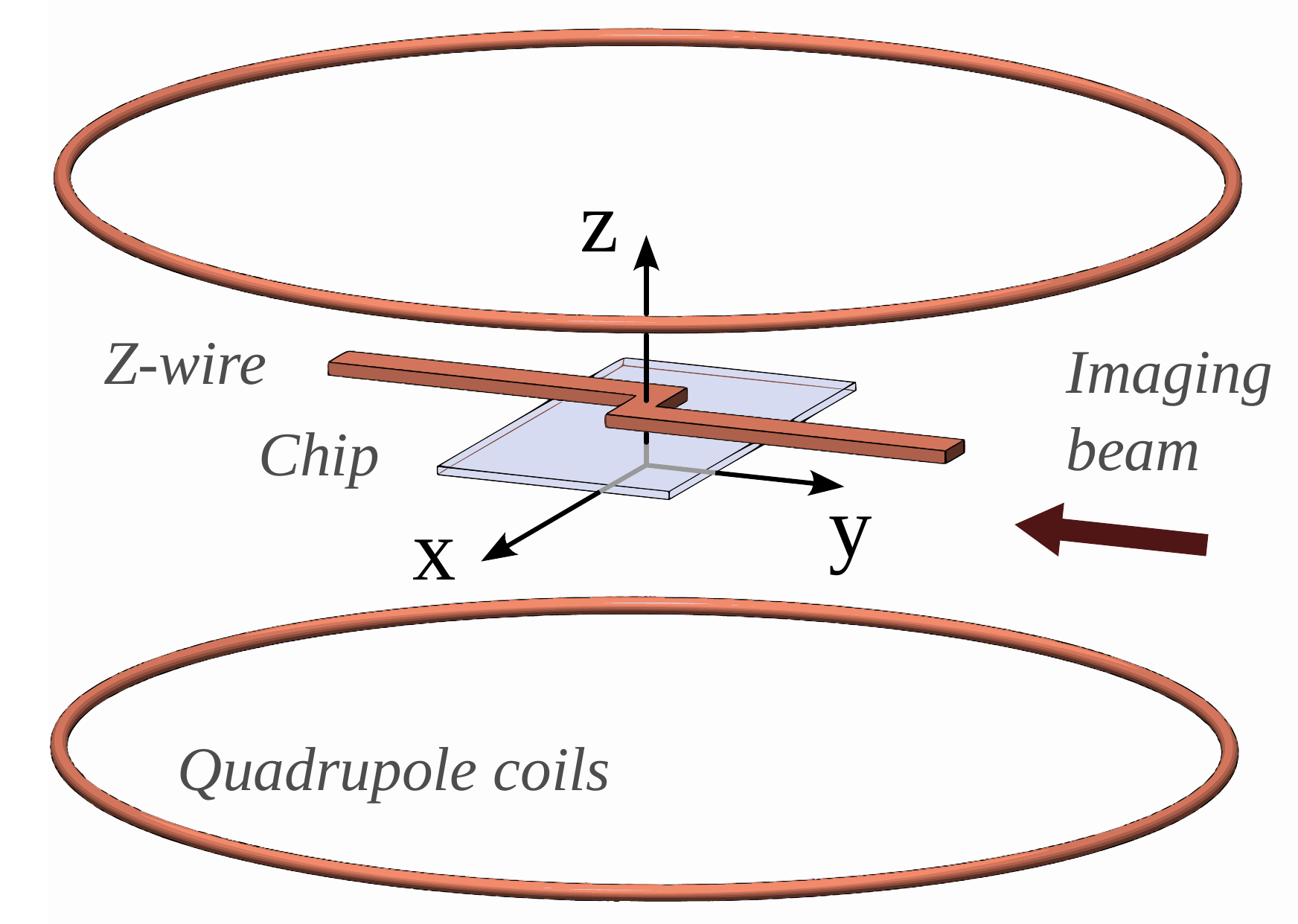}
    \caption{Trapping elements layout and orientation. The on-chip trapping wires (with current flowing along the $x$ direction) are located \SI{4}{\milli\metre} above the zero point of the quadrupole field. The Z-wire that creates the intermediate trap lies above the chip (see Fig.~\ref{fig:layoutnZ} for details). Coils generating homogeneous magnetic fields are not shown. Absorption imaging is performed with a pulse of a resonant laser beam aligned along $-y$.
    \label{fig:traps_and_axes}
    }
    \end{figure}

\begin{figure}[t]
    \centering
    \includegraphics[width=\linewidth]{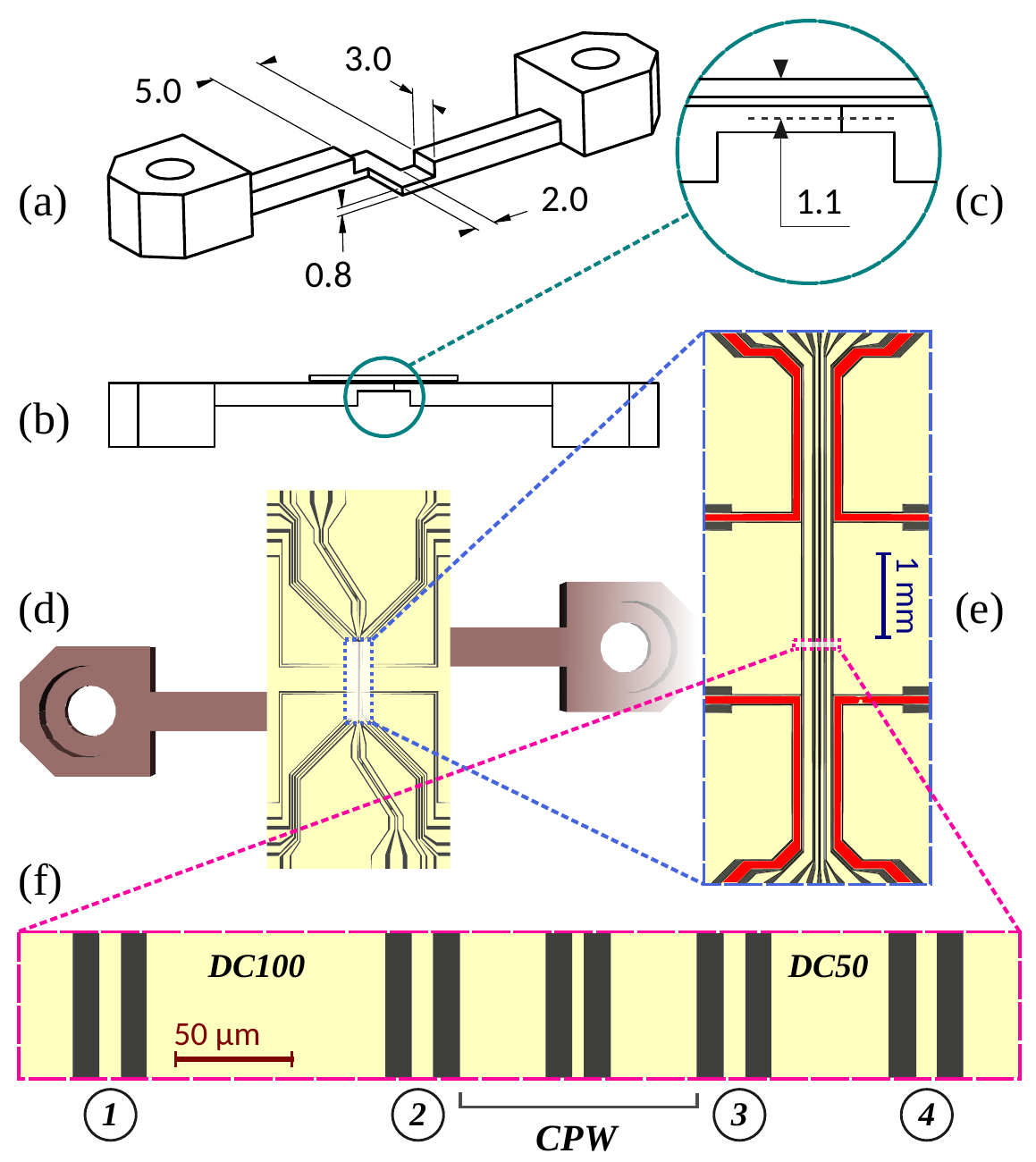}
    
    \caption{\textbf{(a)} Drawing of the Z-wire for the intermediate trap. The dimensions of the central part are given in millimeters. \textbf{(b)} Position of the chip relative to the Z-wire. \textbf{(c)} Zoom of sketch (b). The distance between the top surface of the chip and the median plane of the Z-wire is \SI{1.1}{\milli\metre}. \textbf{(d)} Perspective top-view of the assembly. \textbf{(e)} Close-up view of the chip. For clarity, the four U-shaped wires are colored in red. \textbf{(f)} Zoom in on the central part of the chip. Gold wires in yellow, insulating gaps in black.
    The coplanar waveguide (CPW) consists of three conductors (37, 6 and \SI{37}{\micro\metre} wide respectively) separated by \SI{10}{\micro\metre} gaps. The two trapping wires `DC100' and `DC50' are \SI{100}{\micro\metre} and \SI{50}{\micro\metre} wide respectively.
    The four tracks marked by circled numbers are \SI{10}{\micro\metre} wide and are not used in this work.}
    \label{fig:layoutnZ}
\end{figure}

The overall layout of our experimental configuration is depicted in Fig.~\ref{fig:traps_and_axes}. During the experimental sequence, three different magnetic traps are subsequently used to confine the atoms (see details in Appendix~\ref{sec:sequence}). Two coils of axis $z$ with currents flowing in opposite directions produce a spherical quadrupole magnetic trap. The atom chip lies in the horizontal plane approximately \SI{4}{\milli\metre} above the center of the quadrupole trap, facing downwards. It is made of a silicon wafer of thickness \SI{500}{\micro\metre}, oxidized on a layer of thickness \SI{20}{\nano\metre}. On this face, \SI{2}{\micro\metre}-thick microfabricated wires have been evaporated (see Fig.~\ref{fig:layoutnZ}(d-f)). Two wires are colinear with the $x$-axis and will be referred to as `main trapping wires' in the following: `DC50' and `DC100' with respective transverse widths \SI{50}{\micro\metre} and \SI{100}{\micro\metre}, see Fig.~\ref{fig:layoutnZ}(f). Two pairs of U-shaped wires, or U-wires, of transverse width \SI{100}{\micro\metre} and spaced by \SI{2}{\milli\metre} along the $x$-axis (in red in Fig.~\ref{fig:layoutnZ}(e)) sit on both sides. In between the two main trapping wires, three wires colinear with the $x$-axis and of respective widths \SI{37}{\micro\metre}, \SI{6}{\micro\metre} and \SI{37}{\micro\metre} form a coplanar waveguide that induces a near-field microwave radiation. The geometry of the waveguide has been optimized for a good impedance matching between 1.5 and \SI{2}{\giga\hertz} (see Appendix~\ref{sec:cpw_S_pars}).

Setting currents in one of the main trapping wires as well as in the U-wires and combining with homogeneous external bias fields produced by macroscopic coils surrounding the vacuum chamber, see Appendix~\ref{sec:calib_mag_field} for details, leads to very elongated Ioffe-Pritchard magnetic traps sitting below the atom chip plane~\cite{Reichel2002, Folman2002}. The external bias field has both a transverse component, $\mathbf{B}_{\textrm{bias},\perp}=B_{\textrm{bias},y}\mathbf{e}_y+B_{\textrm{bias},z}\mathbf{e}_z$, whose modulus $B_{\textrm{bias},\perp}$ tunes the distance of the trap minimum to the main wire while the ratio $B_{\textrm{bias},z}/B_{\textrm{bias},y}$ tunes its angular position, and a longitudinal component $B_{\textrm{bias},x}\mathbf{e}_x$ which controls the value of the magnetic field at the trap minimum $B_{\rm min}$.

This geometry results in an almost isotropic harmonic trapping in the transverse plane $yz$ for low-field seeking atomic states. The longitudinal harmonic trapping is controlled independently from the current in the U-wires, that provides a weak curvature to the magnetic potential in the $x$ direction. The longitudinal and transverse oscillation frequencies obtained from this geometry for the magnetic trap are very different, with a longitudinal frequency $\omega_x/(2\pi)$ between 10 and \SI{25}{\hertz} and a transverse frequency $\omega_\perp/(2\pi)$ ranging between 1 and \SI{10}{\kilo\hertz}, depending on the parameters.

A millimeter-sized Z-shaped wire, or Z-wire, is placed \SI{200}{\micro\metre} above the top surface of the atom chip, see Figs.~\ref{fig:traps_and_axes} and \ref{fig:layoutnZ}(a-d). The central bar of this Z-wire plays the role of the main trapping wire and its two side bars the role of longitudinal trapping for a `Z-trap' used as an intermediate step between the macroscopic quadrupole trap and the final microtrap, which have a very different trapping volume (see Appendix~\ref{sec:sequence}).

\section{Degenerate Bose gas in the 3D-1D crossover}
\label{sec:bec}
In this section, we briefly describe the steps taken to produce a Bose-Einstein condensate in the chip trap and examine the dimensionality regime depending on the atom number and temperature.

\subsection{Overview of the experimental steps to degeneracy}

Sodium atoms are loaded in a magneto-optical trap from a permanent-magnet Zeeman slower \cite{Cheiney2011,Ali2017}.
The cold gas is then compressed and further cooled in an optical molasses. A quadrupole magnetic trap is subsequently switched on to capture atoms in the $|F=1,m_F=-1\rangle$ state, where $F$ is the total angular momentum of an atom in its ground state and $m_F$ its projection along the local magnetic-field direction. Taking advantage of a magnetic transport with a sequence of 13 pairs of coils \cite{Greiner2001,Badr2019}, we move the trapped gas over \SI{60}{\centi\metre} up to the science chamber. At the end of the procedure, typically $10^9$ atoms are confined at a temperature of \SI{130}{\micro\kelvin} in the spherical quadrupole trap below the atom chip (see Fig.~\ref{fig:traps_and_axes} for the layout).

The route from this initial trap to the evaporation to degeneracy in the atom chip trap proceeds in two main steps.
The first step brings the atoms from the quadrupole trap to the intermediate Z-trap, for a better mode matching to the tight and elongated final microtrap. A first evaporation ramp is performed in this intermediate trap. The microtrap is loaded during the second step where cooling to degeneracy is performed. Appendix~\ref{sec:sequence} gives further details on the experimental sequence, together with the evolution of atom number, temperature and phase-space density during the two evaporation steps. In the end, we obtain condensates with typically $10^6$ atoms at a temperature of 1~$\mu$K.

\subsection{3D-1D crossover}

Adjusting the final frequency of the forced evaporative cooling ramp allows us to decrease the temperature of the system even further. At some point the thermal fraction becomes negligible and in order to evaluate the temperature $T$ of the gas, we cannot rely on the usual time-of-flight technique. Instead, we take advantage of the very elongated nature of the system: it is responsible for visible density fluctuations along the long direction of the trap that develop during time-of-flight expansion~\cite{dettmer:01,jo:07b,Shah2023}. The power spectrum of these fluctuations contains information on the temperature of the system~\cite{imambekov:09,Schemmer2018}. These effects become significant as soon as the energy scales of the gas, chemical potential $\mu$ and $k_BT$, approach the transverse trapping energy $\hbar\omega_\perp$, where $\hbar$ is the reduced Planck constant and $k_B$ is the Boltzmann constant. In this case, the system enters the 3D-1D crossover, where the physics is mainly governed by the longitudinal characteristics of the gas. The gas cannot be considered as a true Bose-Einstein condensate anymore but is rather referred to as a quasicondensate, where longitudinal thermal excitations population is significant compared with the true ground state population.

\begin{figure}[t]
     \centering \includegraphics{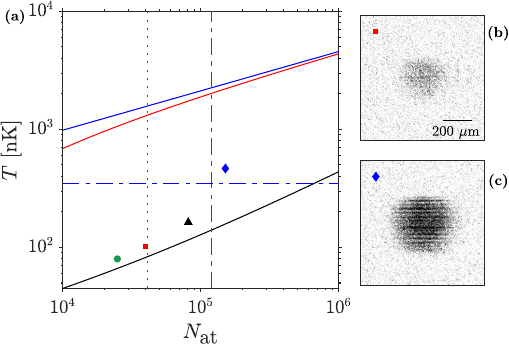}
   \caption{\label{1d3d} \textbf{(a)} Temperature and atom number of the system in the degenerate regime for different final frequencies of the evaporative cooling ramp (blue diamond, black triangle, red square and green disk). The blue line corresponds to the usual three-dimensional temperature threshold for Bose-Einstein condensation $k_BT_c=0.94\,\hbar\overline{\omega}N^{1/3}$ \cite{Dalfovo1999} which assumes $k_B T_c\gg \hbar\overline{\omega}$ with $\overline{\omega}=(\omega_x\omega_\perp^2)^{1/3}$. The red line gives the exact calculation for $T_c$ in our elongated geometry where the semiclassical approximation does not always hold. The red dashed dotted line corresponds to the limit $\chi=\chi_\textrm{cross}$ (see text) and the blue dashed dotted line to $k_BT=\hbar\omega_\perp$. The red dotted line corresponds to the limit $\mu=2\hbar\omega_\perp$. Finally, the black line corresponds to the limit where the thermal coherence length of the gas is equal to the longitudinal radius of the system. \textbf{(b)} and \textbf{(c)} show typical examples of the gas after \SI{10}{\milli\second}  time of flight and obtained with absorption imaging. They correspond to the red square and blue diamond data of \textbf{(a)} respectively.
}
\end{figure}

To evaluate in which regime of the 3D-1D crossover the gas is prepared, we should compare $\mu$ and $k_BT$ to $\hbar\omega_\perp$. Let us first examine the criterion involving $\mu$. To this aim, we use the parameter $\chi=N a a_\perp/a_x^2$ introduced in Ref.~\cite{gerbier:04c}, where $a$ is the scattering length describing the two-body interaction between the atoms and $a_{x,\perp}=\sqrt{\hbar/m\omega_{x,\perp}}$ are the harmonic-oscillator lengths along the respective trap axes, with $m$ the atom mass. According to Ref.~\cite{gerbier:04c}, the boundary between the 3D and 1D regimes occurs for $\chi=\chi_\textrm{cross}\simeq3.73$. Another criterion is obtained by comparing directly the chemical potential $\mu$ to $\hbar\omega_\perp$. $\mu$ is always larger than $\hbar\omega_\perp$ due to the contribution of the zero-point energy $\hbar\omega_\perp$ of the transverse ground state. The cross-over thus occurs when $\mu$ is of order $2\hbar\omega_\perp$.
The chemical potential $\mu$ can be evaluated by solving \cite{gerbier:04c}
\begin{align}
8\left(\frac{\mu}{\hbar\omega_\perp}-1\right)^3\left(\frac{2}{3}\frac{\mu}{\hbar\omega_\perp}+1\right)^2=\left(5\chi\right)^2 .   
\end{align}
The boundary between the 1D and 3D regimes translates into a threshold for the atom number $N$ which is represented in Fig.~\ref{1d3d} using these two criteria.

We now examine the criterion on the temperature. The 1D regime is reached when $k_BT<\hbar\omega_\perp$. As mentioned above, $T$ can be extracted from the analysis of the power spectrum of the density fluctuations that arise during the time of flight expansion. To this aim, we repeat the experiment about 50 times in the same conditions in terms of atom number and final frequency of the evaporative cooling ramp and let the atoms expand for \SI{10}{\milli\second}. The analysis of the pictures recorded by absorption imaging gives access to the power spectrum of the density fluctuations that we compare with the analytical model detailed in~\cite{Schemmer2018}. It allows us to extract the thermal coherence length of the gas $\ell_c=2\hbar^2 n_0/(m k_B T)$, where $n_0$ is the peak density of the longitudinal profile integrated along the transverse directionsnearly equal to its value in the trap as the expansion along the longitudinal direction $x$ during the time of flight is very small \cite{castin:96}. 

In Fig.~\ref{1d3d}, we show the results obtained for four different values of the final frequency at the end of the evaporative cooling ramp, in a trap where $\omega_x=2\pi\times\SI{20}{\hertz}$ and $\omega_\perp=2\pi\times\SI{7.3}{\kilo\hertz}$. Comparing the atom number to the one set by $\chi_\textrm{cross}$ and the temperature to $\hbar\omega_\perp/k_B$, we observe that our system smoothly crosses the 3D-1D crossover. For the two coldest samples, the estimated thermal coherence length $\ell_c$ gets very close to the longitudinal radius of the gas $R_x=a_x^2/a_\perp\sqrt{2(\mu/\hbar\omega_\perp-1)}$ (see~\cite{gerbier:04c}). This means that the system crosses over to a finite-size one-dimensional Bose-Einstein condensate, where the ground-state population becomes preponderant compared with thermal excitations.

\section{Observation of fast coherent Rabi oscillations}\label{fast_rabi}
Once the degenerate gas is prepared in the chip trap, we can manipulate its internal state using the strong microwave field produced in the vicinity of the microwave guide.

\subsection{Moving the cloud below the microwave waveguide}\label{sec:moving}

The quantum gas is prepared at the vertical of the DC100-wire. On the other hand, the microwave waveguide produces an oscillating magnetic field whose amplitude decreases with the distance to the center of the waveguide. To benefit from large microwave amplitudes, we need to move the center of the atom chip trap closer to the waveguide. For this, we change the ratio $B_{\textrm{bias},z}/B_{\textrm{bias},y}$ in order to rotate the position of the center of the trap around the main trapping wire.

\begin{figure}[t]
     \centering \includegraphics{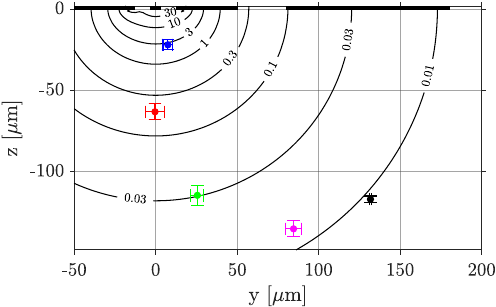}
   \caption{\label{position} Position of the center of the atom chip trap in the $yz$~plane for different values of $\mathbf{B}_{\textrm{bias},\perp}$ : the black point corresponds to the position of the atoms after the atom chip trap loading. The blue point is the closest position to the microwave waveguide explored in this work. Magenta, green and red points are in-between positions. On the upper edge of the figure, the main trapping wire DC100 and the three parts of the microwave waveguide are represented as black rectangles. The black lines correspond to the microwave field isomagnetic lines as deduced from a static model for the waveguide (see text). The modulus of the microwave magnetic field is indicated on each line (in gauss).
}
\end{figure}

From the position where we have reached degeneracy, we linearly increase $B_{\textrm{bias},z}$ while decreasing $B_{\textrm{bias},y}$ in \SI{100}{\milli\second}, just before the final evaporative cooling ramp. In Fig.~\ref{position}, we show different final positions that can be reached depending on the final components of $\mathbf{B}_{\textrm{bias},\perp}$. We also indicate the value of the amplitude of the microwave field that can be deduced from a simple model which assumes static currents in the waveguide, as follows. The skin depth in a gold wire for a signal around \SI{1.77}{\giga\hertz} is about \SI{1.8}{\micro\metre}, and we assume a homogeneous current density in the central wire. For the two ground wires, the current flows in the opposite direction with half the amplitude and we assume a homogeneous current density at the two inner edges facing the central wire, spread over a width set by the skin depth.

The positions shown in Fig.~\ref{position} are deduced from a model which takes into account the geometry of the DC100- and U-wires assuming a homogeneous current density. The two components of $\mathbf{B}_{\textrm{bias},\perp}$ have been calibrated independently by microwave spectroscopy (see Appendix~\ref{sec:calib_mag_field}).

\subsection{Calibration of the microwave field amplitude}\label{sec:calibmw}

\begin{figure}[t]
     \centering \includegraphics{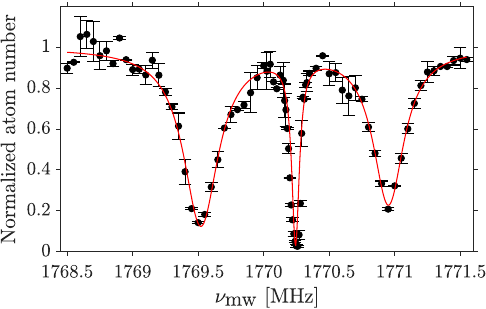}
   \caption{\label{three_pics} Microwave spectroscopy of the trapped atoms (see Appendix~\ref{sec:calib_mag_field} for the exact experimental procedure). The three peaks correspond to the different polarizations present in the microwave field. The distance between the peaks is used to determine $B_{\rm min}$ precisely as described in Appendix~\ref{sec:calib_mag_field}. Here we find $B_{\rm min}=1.02$\,G. The red line is a fit including three Lorentzian peaks.
}
\end{figure}

A more precise calibration of the microwave field amplitude with respect to the simple model introduced in Sec.~\ref{sec:moving} is performed by a direct measurement of the field amplitude using the atomic cloud as a local probe. This also gives access to the amplitude of the different components of the microwave field polarization. To this aim, we drive Rabi oscillations between the trapped $|F=1,m_F=-1\rangle$ state and one of the $|F=2,m_F=-2,-1,0\rangle$ states that are accessible due to selection rules. Figure~\ref{three_pics} shows a spectrum where these three lines are visible, evincing different coupling strengths. To observe clear oscillations of the population in the initial state, we need to choose a Rabi frequency $\Omega_R$
larger than the chemical potential of the trapped gas, to limit the loss of coherence due to the inhomogeneous magnetic field in the trap, see Sec.~\ref{sec:damping}, but smaller than the Zeeman splitting to resolve a single two-level transition.
The expression of the Rabi frequency $\Omega_R$ for a given isolated hyperfine transition is detailed in Appendix~\ref{sec:mwcoupling}. Its relation to the microwave amplitudes $\Omega_{\pm,0}=-|g_F|\mu_BB_{\pm,0}/\hbar$ in units of frequency, where $B_\pm$ and $B_0$ are the microwave field polarization components, depends on the matrix elements of the transition. Briefly, $\Omega_R$ is equal to the microwave amplitude $\Omega_+$ for the $|F=1,m_F=-1\rangle\to|F=2,m_F=0\rangle$ transition addressed by the $\sigma^+$ component of the polarization, to $\sqrt{3}\Omega_0$ for the $|F=1,m_F=-1\rangle\to|F=2,m_F=-1\rangle$ $\pi$ transition and to $\sqrt{6}\Omega_-$ for the $|F=1,m_F=-1\rangle\to|F=2,m_F=-2\rangle$ $\sigma^-$ transition. The chemical potential is $h\times\SI{20}{\kilo\hertz}$ or below for the typical atom numbers and trap geometries that we have investigated. The Zeeman frequency splitting is on the order of $\SI{700}{\kilo\hertz}$ at the trap bottom, such that we aim for Rabi frequencies on the order of $\Omega_R=2\pi\times\SI{100}{\kilo\hertz}$.

\begin{figure}[t]
     \centering \includegraphics{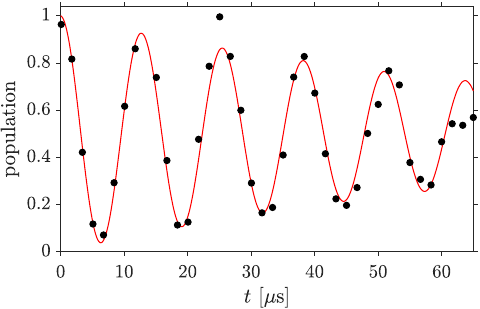}
   \caption{\label{fig:rabi_low_power} Rabi oscillations of the population of the $|F=1,m_F=-1\rangle$ state coupled to the $|F=2,m_F=-2\rangle$ state with the $\sigma^-$ component of the microwave field. From the fit of the Rabi frequency $\Omega_R$, which includes an exponential damping of the oscillations, we deduce a microwave amplitude $|\Omega_-|=\Omega_R/\sqrt{6}\simeq2\pi\times\SI{32}{\kilo\hertz}$. The different contributions to the damping of the oscillations are discussed in the text.
}
\end{figure}

At each trap position shown in Fig.~\ref{position}, we adjust the amplitude of the current in the microwave waveguide in order to observe clear oscillations of the population of the trapped $|F=1,m_F=-1\rangle$ state while scanning the microwave pulse duration. A typical example is shown in Fig.~\ref{fig:rabi_low_power}. We keep the atoms in the trap for \SI{20}{\milli\second} after the pulse in order to get rid of the $F=2$ population which is not trapped. We then fit the frequency of the oscillations in order to determine the Rabi frequency $\Omega_R$, and thus the amplitude of the component of the microwave field which drives the oscillations. In the case of Fig.~\ref{fig:rabi_low_power}, the microwave is resonant with the $|F=1,m_F=-1\rangle\rightarrow|F=2,m_F=-2\rangle$ transition and we find $|B_-|\simeq46$~mG. We also observe a damping of the oscillations that we attribute to a gradient of amplitude of the microwave field over the vertical size of the cloud and to the inhomogeneity of the magnetic field in the trap, see Sec.~\ref{sec:damping}. 

We have repeated this procedure for all the trap positions shown in Fig.~\ref{position}, calibrating $|B_-|$ and $|B_+|$. Even though we observe a small coupling to $|F=2,m_F=-1\rangle$ due to $|B_0|$ for a long pulse duration (see Fig.~\ref{three_pics}), its contribution is much smaller. This can be understood from the static model for the microwave waveguide introduced in Sec.~\ref{sec:moving}. It predicts that the polarization of the field should be linear and orthogonal to the $x$ axis. 
We still expect a faint contribution from the $\pi$ polarization component of the microwave field with respect to the quantization axis set by the local static field due to two effects.
First, the static field at the center of the trap axis is slightly tilted in the $(xy)$ plane from the $x$ axis by approximately $\SI{18}{\milli\radian}$. Second, the direction of the static field varies over the atomic cloud by up to $\arctan\left[\sqrt{2\mu/(|g_F|\mu_B B_{\rm min})}\right]=\SI{235}{\milli\radian}$ at the edge for a chemical potential of $\mu=h\times\SI{20}{\kilo\hertz}$ and $B_{\rm min}=1$\,G. This last contribution is larger, leading to a small coupling to the $\pi$ transition, limited to a few percent of the coupling to the $\sigma^-$ transition when averaged over the size of the cloud. In addition, it is responsible for an associated spread of the microwave amplitude for the $\sigma^+$ and $\sigma^-$ transition. However, it remains below 3\%, and in the following we will assume that the direction of the static field is uniform over the cloud.

\begin{figure}[t]
     \centering \includegraphics{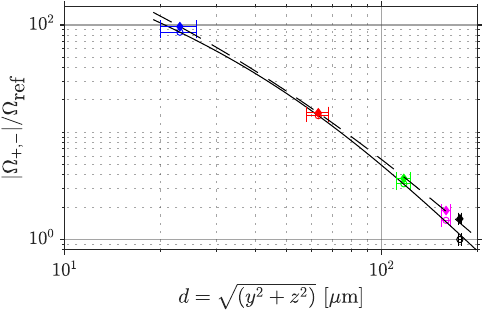}
   \caption{Relative microwave amplitudes $|\Omega_{+}|/\Omega_{\rm ref}$ (filled diamond) and $|\Omega_{-}|/\Omega_{\rm ref}$ (circle) as a function of the distance to the center of the waveguide, normalized by the value $\Omega_{\rm ref}$ of $|\Omega_{-}|$ at the initial position, the farthest from the waveguide. The color code is the same as in Fig.~\ref{position}. The solid (dashed) line is an interpolation of $|\Omega_{-}|/\Omega_{\rm ref}$ ($|\Omega_{+}|/\Omega_{\rm ref}$).
}\label{fig:amplitudeMW}
\end{figure}

To compare the results obtained at different positions, we have rescaled the measured microwave amplitude $|\Omega_{+,-}|$ assuming that it depends linearly on the amplitude of the current in the microwave waveguide. The results are shown in Fig.~\ref{fig:amplitudeMW}, the color code matching the one of Fig.~\ref{position}. From the static model, the decay of the amplitude of the microwave field is expected to scale as $1/d$ at short distance and as $1/d^3$ at long distance. Most of our data belong to a region which interpolates between these two regimes. We also see that the amplitudes of the $\sigma^-$ and $\sigma^+$ components of the microwave field are not exactly equal at a given position. This effect is not captured by the static model of the waveguide. The measured values of $|\Omega_{+,-}|$ before rescaling are given in Table~\ref{tab:raw_data}.

\subsection{Damping at moderate microwave amplitude}
\label{sec:damping}
We now discuss the origin of the damping observed at moderate microwave amplitude, as illustrated in Fig.~\ref{fig:rabi_low_power}. Damping may have several origins such as different interactions in the different hyperfine states, inhomogeneous magnetic frequency shifts, microwave amplitude inhomogeneity, or the effect of external motion during the pulse. Let us estimate the importance of these effects for our parameters.

We first remark that the scattering lengths in the different hyperfine states of $F=1$ and $F=2$ differ by at most 40\% \cite{Tiesinga1996,Crubellier1999,Knoop2011}, leading to a small positive shift of the transition \cite{Harber2002} (a fraction of the chemical potential). As the density is inhomogeneous, the main effect is a small frequency broadening. However this broadening is smaller that the frequency broadening due to the inhomogeneous Zeeman effect, and is further reduced if the cloud expands during the pulse. Its contribution to the damping observed in each of our data is thus minor compared with other sources. 

Frequency broadening due to the Zeeman effect is expected because the initial state $|F=1,m_F=-1\rangle$ is trapped with an energy spread equal to $m\omega_\perp^2R_\perp^2/2$ where $R_\perp$ is the transverse radius. In contrast, the final state is either independent of magnetic field for the $\sigma^+$ transition, or undergoes an opposite magnetic shift for the $|F=2,m_F=-1\rangle$ state, doubled for the $|F=2,m_F=-2\rangle$ state. The transition thus acquires a half-width $\Delta_Z=\kappa\mu/2\hbar$ where $\kappa=1$, 2 or 3 for the $\sigma^+$, $\pi$ and $\sigma^-$ transitions, respectively. If the local detuning due to the Zeeman shift is $\delta$, the off-resonant frequency $(\Omega_R^2+\delta^2)^{1/2}$ of the Rabi oscillation leads to a broadening of the oscillation frequency in the gas scaling as $\Delta_Z^2/\Omega_R$ for $\Delta_Z<\Omega_R$. This results in a damping time of order $\tau_Z=2\pi\Omega_R/\Delta_Z^2$, corresponding to a number of Rabi oscillation periods of $\Omega_R\tau_Z/(2\pi)=[\Omega_R/\Delta_Z]^2$. For $\Delta_Z>\Omega_R$, Rabi oscillations are suppressed.

An amplitude broadening is also present, due to a gradient of microwave amplitude over the radial size of the atomic cloud $2R_\perp$, more pronounced for the positions closest to the waveguide. The relative microwave amplitude difference $\eta$ between the two edges of the gas when it is initially trapped is about $\eta=3\%$ at the position furthest from the waveguide, and increases up to about $\eta=12\%$ at the position closest to the waveguide that we have investigated, as estimated from the simple static model of the microwave field, Sec.~\ref{sec:moving}. The time constant for damping associated with this effect is $\tau_\Omega\approx 2\pi/\Delta_\Omega$ where $\Delta_\Omega=\eta\Omega_R$ is the difference in Rabi frequency between the two edges of the cloud. It sets a maximum number of Rabi periods $\Omega_R\tau_\Omega/(2\pi)=\eta^{-1}$ that can be observed, ranging between 8 near the waveguide to 30 at the largest distance probed here.

The comparison of these two timescales $\tau_Z$ and $\tau_\Omega$ shows that frequency broadening will set the shortest damping time at small amplitude, while amplitude gradient will set the limit at large amplitude, namely for $\Omega_R>\Delta_Z/\sqrt{\eta}$. We note that in our experiment, $\Delta_Z/\sqrt{\eta}$ is always at least five times larger that $\omega_\perp$ such that in the regime where amplitude gradient effects dominate the damping, the transverse motion can be neglected.

In contrast, the damping mechanism due to frequency broadening may be affected by the external motion in the transverse direction. For the three transitions investigated in this paper, the final sublevel in $F=2$ hyperfine state is not trapped. The external degrees of freedom will thus evolve, with a time constant set by the oscillation frequencies $\omega_x^{-1}\simeq \SI{10}{\milli\second}$ and $\omega_\perp^{-1}\simeq \SI{40}{\micro\second}$. While the longitudinal motion is very slow with respect to the Rabi oscillations observed in this work, the radial motion has a timescale comparable to the pulse duration at moderate microwave Rabi frequency $\Omega_R/(2\pi)\leq\SI{200}{\kilo\hertz}$. As a consequence, the cloud radius $R_\perp$ can expand or oscillate radially during the pulse. 
If $R_\perp$ increases by a factor $\lambda_\perp$, the Zeeman width $\Delta_Z$ is increased by $\lambda_\perp^2$ and the damping time $\tau_Z$ reduced by a factor $\lambda_\perp^4$.

We evaluate this expansion in a Castin-Dum approach \cite{castin:96}, assuming that it results in a time-dependent transverse scaling factor $\lambda_\perp(t)$ for the transverse radius $R_\perp$ while $R_x$ is unchanged. We assume that the atoms undergo an average potential resulting from equal population in the initial and final state, and solve the Castin-Dum equations for $\lambda_\perp$:
\begin{equation}
    \ddot{\lambda}_\perp = \frac{\omega_\perp^2}{\lambda_\perp^3} + \frac{\kappa-2}{2}\omega_\perp^2\lambda_\perp.
\end{equation}
This estimation leads to (i) an exponential growth for the $\sigma^-$ transition where the average potential is repulsive ($\kappa=3$), with 
$\lambda_\perp(t)=\left[1+3\sinh^2(\omega_\perp t/\sqrt{2})\right]^{1/2}$,
(ii) a scaling
$\lambda_\perp(t)=\left[1+\omega_\perp^2t^2\right]^{1/2}$ 
similar to a time-of-flight expansion for the $\pi$ transition where the average potential vanishes ($\kappa=2$), and (iii) a scaling
$\lambda_\perp(t)=\left[1+\sin^2(\omega_\perp t/\sqrt{2})\right]^{1/2}$ 
corresponding to a transverse monopole excitation for the $\sigma^+$ transition where the average potential is an harmonic trap with a reduced frequency ($\kappa=1$). In this last case, the effect of the transverse motion is limited as the transverse radius increases by at most a factor $\sqrt{2}$, reducing $\tau_Z$ by at most a factor 4.

For the two other cases where the transverse radius expands, we can estimate the damping time set by the Zeeman effect as the time $\tau'_Z$ after which the frequency broadening $\Delta_Z\lambda_\perp$ reaches $\Omega_R$. It turns out that in our case, this time is always shorter than $\tau_Z$ or of the same order. We find $\tau'_Z=\omega_\perp^{-1}\sqrt{\hbar\Omega_R/\mu-1}$ for the $\pi$ transition and $\tau'_Z=\sqrt{2}\omega_\perp^{-1}\mbox{arcsinh}\sqrt{\left(2\hbar\Omega_R/(3\mu)-1\right)/3}$ for the $\sigma^-$ transition. For the data presented in Fig.~\ref{fig:rabi_low_power} this estimation yields $\tau'_Z\approx\SI{50}{\micro\second}$, a bit shorter than the observed damping time of $\SI{75}{\micro\second}$.

\begin{table}[t]
    \centering
    \begin{tabular}{c | c | c c | c c}
        \hline\hline
        &&&&&\\[-3mm]
    $d$ & Power & $|\Omega_-|/2\pi$ & $|B_-|$ & $|\Omega_+|/2\pi$ & $|B_+|$\\
    ~[\SI{}{\micro\metre}]~ & [dBm] & [kHz] & [mG] & [kHz] & [mG]\\
    \hline\hline
       \multirow{2}{1.5em}{176} & 36 & 43 & 61 & 67 & 96\\
         & 32 & 27 & 39 & &\\
         \hline
         \multirow{2}{1.5em}{159} & 32 & 40 & 57 & 51 & 73 \\
         & 23  & 15 & 21 & &\\
         \hline
        117 & 23 & 32 & 46 & 36 & 51\\
         \hline
         \multirow{2}{1.5em}{63} & 23 & 137 & 200 & 146 & 210 \\
         & 18 & 78 & 110 & &\\
         \hline
         \multirow{2}{1.5em}{23} & 18 & 461 & 660 & &\\
         & 13 & 262 & 370  & 294 & 420\\
             \hline\hline
    \end{tabular}
    \caption{
Raw data used to produce Fig.~\ref{fig:amplitudeMW}. The distance $d$ of the trap position to the main trapping wire is computed assuming a homogeneous current density in the atom chip wires. The microwave power is measured at the input of the circuit of the coplanar waveguide (see Appendix~\ref{sec:cpw_S_pars} for its spectral response).
The amplitude $|\Omega_-|$ and $|\Omega_+|$ and microwave amplitudes $|B_-|$ and $|B_+|$ are deduced from a fit of the population oscillations between the trapped $|F=1,m_F=-1\rangle$ state and the $|F=2,m_F=-2\rangle$ and $|F=2,m_F=0\rangle$ respectively (see for instance Fig.~\ref{fig:rabi_low_power}).
The microwave frequency is adapted to be resonant with the probed transition.
    \label{tab:raw_data}}
\end{table}

\subsection{Large amplitude coherent oscillations}\label{sec:large}

\begin{figure}[t]
     \centering \includegraphics{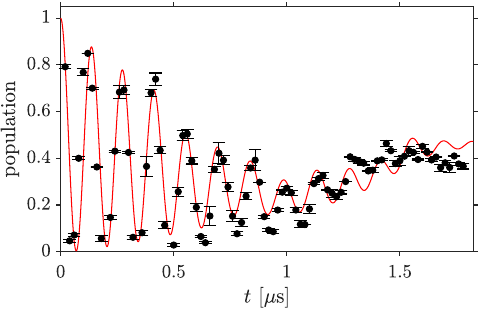}
   \caption{\label{fig:rabi_high_power} Evolution of the population of the $|F=1,m_F=-1\rangle$ trapped state coupled to the $F=2$ states with a large amplitude microwave field resonant with the $|F=1,m_F=-1\rangle\rightarrow|F=2,m_F=-2\rangle$ transition. The red line corresponds to theoretical expectations for a static field of 1\,G and a linearly polarized microwave field with $|\Omega_{-}|=|\Omega_{+}|\simeq2\pi\times\SI{2.8}{\mega\hertz}$. For the two levels coupled by the $\sigma^-$ transition, this corresponds to a Rabi frequency $\Omega_R=2\pi\times\SI{6.9}{\mega\hertz}$. The damping is captured by a Gaussian distribution of microwave amplitudes $|\Omega_{+,-}|$ of $1/\sqrt{e}$ half-width $2\pi\times 92$~kHz.
}
\end{figure}

We have studied the evolution of the hyperfine state population in the case where the trap is brought at its closest position to the waveguide (blue point in Fig.~\ref{position}), using a strong microwave amplitude.
With these large amplitudes, we observe rapid coherent oscillations of the population of the trapped $|F=1,m_F=-1\rangle$ state as shown in Fig.~\ref{fig:rabi_high_power}. In this case, both the $\sigma^+$ and $\sigma^-$ polarization contribute since the amplitudes $|\Omega_{+,-}|$ are larger than the Zeeman splitting. Five levels in W configuration then contribute to the shape of the coherent signal, see Appendix~\ref{sec:mwcoupling}. With these very large amplitudes and fast oscillation, the expansion due to the repulsive potential in the upper hyperfine state is negligible, and damping is dominated by the amplitude gradient in the direction normal to the waveguide, as discussed in Sec.~\ref{sec:damping}. At the closest position to the waveguide where $\eta=12\%$, the simple estimate considering only the $\sigma^-$ transition yields $\tau_\Omega=\SI{1.2}{\micro\second}$, compatible with the data presented in Fig.~\ref{fig:rabi_high_power}. 

From the model described in Appendix~\ref{sec:mwcoupling}, we can compute the evolution of the population in the initial state. The amplitude gradient can be taken into account by averaging the oscillating populations over a Gaussian distribution of microwave amplitude. Assuming a perfect linear polarization $B_{\rm mw}\cos(\omega t)\mathbf{e}_y$, and adjusting to the experimental data of Fig.~\ref{fig:rabi_high_power}, we deduce $|\Omega_{-}|=|\Omega_{+}|\simeq2\pi\times\SI{2.8}{\mega\hertz}$ for the center of microwave amplitude distribution. This corresponds to $|B_-|=|B_+|\simeq4$\,G or equivalently $B_{\rm mw}=5.7$\,G. Hence, the Rabi coupling on the $|F=1,m_F=-1\rangle\rightarrow|F=2,m_F=-2\rangle$ $\sigma^-$ transition is $\Omega_R=2\pi\times\SI{6.9}{\mega\hertz}$ while it is $2\pi\times \SI{2.8}{\mega\hertz}$ for the $|F=1,m_F=-1\rangle\rightarrow|F=2,m_F=0\rangle$ $\sigma^+$ transition. Such a large field enables a fast manipulation of the internal state, as the spin can be flipped from $F=1$ to $F=2$ in \SI{72}{\nano\second} only. The fit also leads to a $1/\sqrt{e}$ half-width of the microwave amplitude $|\Omega_{+,-}|$ distribution of $2\pi\times92$~kHz. This corresponds to $\eta=16\%$ which is a bit larger than our rough estimation, but also includes the contribution from the spread in static field orientation discussed in Sec.~\ref{sec:calibmw} as well as possible microwave field inhomogeneity along the $x$ axis.

\section{Conclusion}

In this paper we have described the production of degenerate quantum gases of sodium atoms in a magnetic microtrap relying on an atom chip. The latter encompasses a coplanar waveguide that we use to manipulate the hyperfine state of the trapped atoms thanks to a large amplitude microwave field. We observe coherent Rabi oscillations of the atomic population that allow us to determine the corresponding Rabi frequency and hence the amplitude of the microwave field. Depending on the distance of the magnetic trap minimum to the coplanar waveguide and on the microwave power used, we can tune the Rabi coupling on the strongest transition from a few tens of kHz up to \SI{6.9}{\mega\hertz}.

\begin{figure*}[t]
      \centering\includegraphics{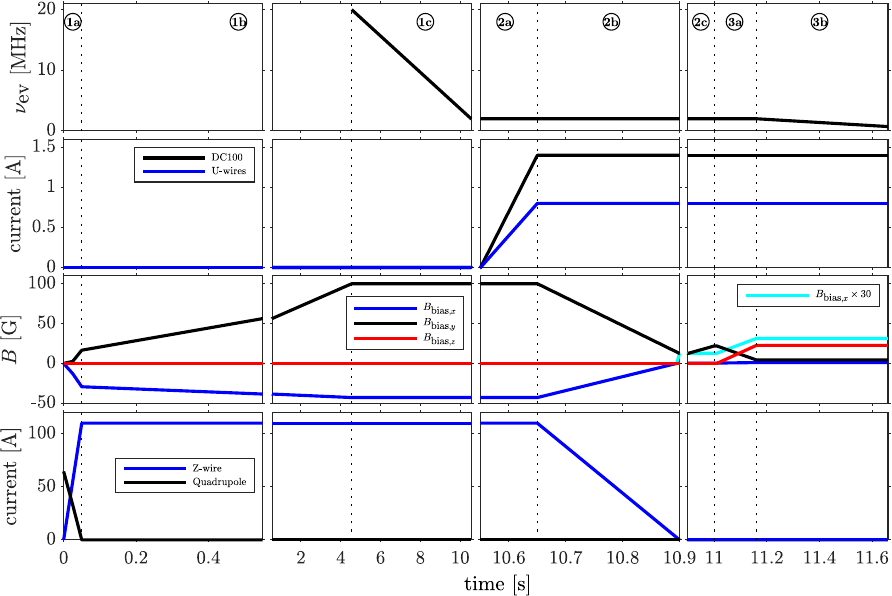}
   \caption{\label{sequence} Timeline of the experimental sequence. From bottom to top, currents in the Z-wire and the quadrupole coils, $x$, $y$ and $z$ components of $B_\textrm{bias}$, currents in the main trapping wire and U-wires and frequency $\nu_\textrm{ev}$ of the evaporative cooling ramp. The sequence is split into three main steps, each divided into substeps: \textbf{1a} Transfer from the quadrupole trap to the intermediate Z-trap - \textbf{1b} Compression of the Z-trap - \textbf{1c} Z-trap evaporative cooling ramp - \textbf{2a} Switch on of the main trapping wire and U-wires current  - \textbf{2b} Transfer to the atom chip trap - \textbf{2c} Compression of the atom chip trap - \textbf{3a} Transport to below the microwave waveguide - \textbf{3b} Final evaporative cooling ramp to degeneracy.
}
\end{figure*}

Such a large coupling can be interesting for quantum technology applications where fast manipulation of the atomic spin states are needed. In this work, $F=2$ spin states that we can couple to with the microwave field are untrapped in the magnetic confinement. It is nevertheless possible to use a two-photon transition, relying on an additional radio-frequency source, to couple to $\vert F=2,m_F=1\rangle$, which is trapped. This state has also the advantage, for a static field of 0.68~G, to experience the same second-order Zeeman shift than the initial trapped state $\vert F=1,m_F=-1\rangle$, allowing for long coherence times~\cite{Treutlein2004}. To avoid coupling to other untrapped states, one should limit the two-photon microwave couplings to below the Zeeman shift between the different spin states, i.e., to about \SI{500}{\kilo\hertz} at this optimum magnetic field.

Large coupling also allows for the realization of adiabatic potentials relying on microwave fields~\cite{Agosta1989} as also demonstrated with radio-frequency fields~\cite{Garraway2016}. In this case, effects beyond the rotating wave approximation should be negligible~\cite{hofferberth:07a}. In the end, the most promising application of this work concerns the possibility to address molecular resonances as explained for alkali atoms in~\cite{Papoular2010} and recently demonstrated for rubidium atoms~\cite{Maury2023}. Close to the resonance, the scattering length of the atoms is expected to be modified as is the case with static magnetic-field Feshbach resonances. Since the width of the resonance depends on the square of the microwave field amplitude, working with large amplitude microwave fields would allow for finer tuning of the scattering length.

\begin{acknowledgments}
We thank Romain Dubessy for the design of the coplanar waveguide and for helpful discussions, Matthias Stüwe and J{\"o}rg Schmiedmayer (TU Wien) for the atom chip fabrication, Thibaut Vacelet (Observatoire de Paris - LERMA (CTOP)) for cutting and performing the wire-bonding on the atom chip. We also thank Jeanne Solard (USPN), Pascal Filloux and Christophe Manquest (USPC) for their assistance in the early stages of the technical development and Yann Charles (LSPM) for performing numerical simulations of the heat conductance of the trapping wires. This work has been supported by the Agence Nationale de la Recherche (ANR) under the projects `ANR-21-CE47-0009-03' and `ANR-22-CE91-0005-01'.
\end{acknowledgments}

\appendix

\section{Experimental sequence}
\label{sec:sequence}

In this Appendix, we give additional details on the experimental sequence for the production of a degenerate gas. Figure~\ref{sequence} summarizes the timeline for the currents, bias fields and evaporative cooling frequency $\nu_\textrm{ev}$ used to bring the atoms from the initial quadrupole trap to the atom chip trap and then close to the coplanar waveguide. The values shown here correspond to the blue position in Fig.~\ref{position}. We give further details on stage (1) and (2) in the next two sections and Fig.~\ref{fig:psd} gives an overview of the atom number, temperature and phase-space density along the route to degeneracy.

\subsection{First step: Transfer to the intermediate Z-trap and pre-cooling stage}\label{sec:firststep}

The `Z-trap' is an elongated Ioffe-Pritchard trap. It relies on the field produced by a current of \SI{110}{\ampere} flowing in the Z-wire combined with a bias field as introduced in Sec.~\ref{at_ch_design}. The $x$-component of the bias field is chosen to fix a minimum field of order 2\,G, preventing Majorana losses.

The transfer from the initial quadrupole trap to the intermediate Z-trap is performed in two parts: a fast transfer to a tighter quadrupole trap resulting from the Z-wire, step (1a) of Fig.~\ref{sequence}, followed by a slower transfer to the Ioffe-Pritchard trap with a nonzero magnetic minimum (1b). Step (1a) is completed in \SI{50}{\milli\second} through a linear increase of the current in the Z-wire from 0 to \SI{110}{\ampere}, while the current of the quadrupole coils is decreased linearly down to zero. Simultaneously, we ramp up $B_{\textrm{bias},x}$ and $B_{\textrm{bias},y}$ in two successive linear ramps of \SI{25}{\milli\second} each.  Near the end of the ramps, the trap is strongly deformed which leads to significant atomic losses.  At the end of this part of the transfer, the center of the trap is close to its initial position and $B_{\rm min}=0$. At this stage we are left with 55\% of the initial atoms.

The second part (1b) is much slower. During the next \SI{500}{\milli\second}, we increase further both $B_{\textrm{bias},x}$ and $B_{\textrm{bias},y}$, which brings the atoms closer to the surface of the chip, enhances the transverse magnetic gradients and sets the magnetic-field minimum $B_{\rm min}$ at about 2\,G, resulting in a Ioffe-Pritchard configuration. At this point, the cloud size is large enough for the most energetic atoms to collide with the atom chip surface. For the next compression stage of \SI{4}{\second}, we continue to move the trap upwards and compress it very slowly, until the collision rate becomes favorable to start a forced evaporative cooling ramp.

For the next \SI{6}{\second} (1c), we perform evaporation in the Z-trap, at a distance of about \SI{550}{\micro\metre} from the atom chip surface. To expel the most energetic atoms from the magnetic trap, we rely on a radio frequency field produced by a coil located just above the atom chip. We start evaporation with a frequency $\nu_{\rm ev}$ of \SI{20}{\mega\hertz} and linearly ramp it down to \SI{3}{\mega\hertz}.

\begin{figure*}[t]
   \begin{minipage}[c]{.46\linewidth}
      \centering\includegraphics{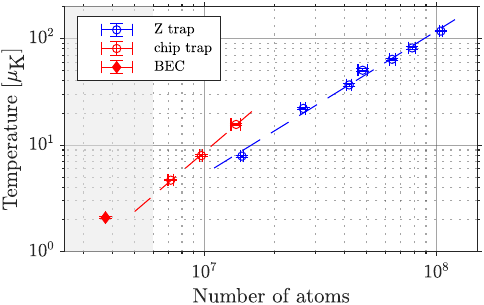}
   \end{minipage} \hfill
   \begin{minipage}[c]{.46\linewidth}
      \centering\includegraphics{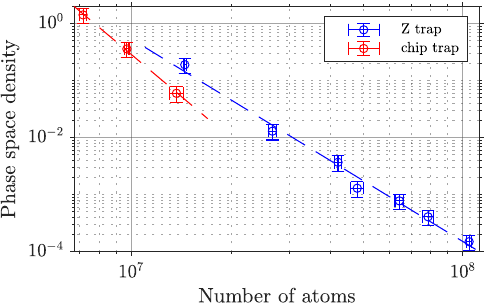}
   \end{minipage}
   \caption{\label{fig:psd} \textbf{Left} - Evolution of the atom number and temperature during the two forced evaporative cooling ramps, respectively in the Z-trap (blue circles) and in the atom chip trap (red circles). The shaded area indicates the region where the system reaches Bose-Einstein condensation. The red filled circle corresponds to a partially degenerate gas and the temperature is evaluated from its thermal fraction. \textbf{Right} - Same results --apart from the last point-- in terms of phase-space density and atom number. The evaluation of the phase-space density relies on a modeling of the trapping potential (see Appendix~\ref{sec:calib_mag_field}).
}
\end{figure*}

The blue circles in Fig.~\ref{fig:psd} show the temperature (left) or phase-space density (right) as a function of the atom number during this evaporation stage. The calibration of atom number is explained in Appendix~\ref{calib_imaging}.
The peak phase-space density is estimated from the measured atom number $N$ and temperature $T$, and from the knowledge of the magnetic landscape (see Appendix~\ref{sec:calib_mag_field}). It is equal to $n(\mathbf{r}_0)\lambda_T^3$, with
\begin{equation}
n(\mathbf{r}_0)= \frac{N \exp{[-V(\mathbf{r}_0)/k_BT]}}{\int d\mathbf{r} \exp{[-V(\mathbf{r})/k_BT]}}
\end{equation}
being the density at the position $\mathbf{r}_0$ of the magnetic-field minimum and 
\begin{equation}
    \lambda_T=\sqrt{\frac{2\pi \hbar^2}{m k_B T}}
\end{equation}
the thermal de Broglie wavelength. The magnetic potential $V$ seen by the trapped atomic state is calculated numerically from the geometry of the different wires involved and relying on an independent calibration of $\mathbf{B}_\textrm{bias}$ explained in Appendix~\ref{sec:calib_mag_field}.

\subsection{Second step: Transfer to the atom chip trap and final evaporation}\label{sndstep}

The transfer from the Z-trap to the atom chip trap is performed in \SI{350}{\milli\second}. We first linearly ramp up the current in the DC100-wire and the U-wires to their final values in \SI{100}{\milli\second} (2a). In the next \SI{250}{\milli\second} we linearly decrease the current in the Z-wire while linearly ramping down $B_{\textrm{bias},x}$ and $B_{\textrm{bias},y}$ (2b). Note that \SI{5}{\milli\second} before the end of the ramp, $B_{\textrm{bias},x}$ vanishes and then changes sign. This procedure allows us to increase the offset field produced by the U-wires along $x$ and maintain the magnetic-field minimum between 1 and 2\,G during the whole transfer. At this point the atoms are confined at a distance of \SI{210}{\micro\metre} from the chip surface.

We then move up and compress the trap further by linearly increasing $B_{\textrm{bias},y}$ to its final value in \SI{100}{\milli\second} (2c). The distance of the atoms to the chip surface is now \SI{120}{\micro\metre}, the measured transverse trapping frequency reaches $\omega_\perp/(2\pi)=\SI{3.2}{\kilo\hertz}$ and the magnetic-field minimum is now $B_{\rm min}=1$\,G. During the transfer and compression stages, we also keep a radio-frequency-knife at a frequency $\nu_{\rm ev}=\SI{3}{\mega\hertz}$. When needed, we then modify the ratio between the bias currents to displace the cloud at the desired distance from the microwave waveguide (3a). During the next \SI{500}{\milli\second} (3b), we proceed to the final forced evaporative cooling ramp from \SI{3}{\mega\hertz} down to \SI{750}{\kilo\hertz} where we reach degeneracy and observe Bose-Einstein condensation. The red circles in Fig.~\ref{fig:psd} summarize our results in term of temperature (left), phase-space density (right) and atom number during this last evaporation stage.

\section{Characterization of the static magnetic landscape}\label{sec:calib_mag_field}

To estimate the phase-space density in the trap or locate the position of the trap center, a good knowledge of the full magnetic potential $V(\mathbf{r})$ and hence of the magnetic landscape is required. To this aim, we look for a relation between the value of the currents in the chip wires, Z-wire and coils, and the resulting static magnetic field. This relation is obtained by an accurate modeling of the wire configuration for the chip wires and Z-wire, and by a direct measurement of the field produced by the macroscopic coils for the bias fields.

Each wire on the atom chip as well as the Z-wire are approximated by a succession of rectangular cuboids with homogeneous current density. The geometry of the cuboids and wires are identical apart close to regions where the current is bent. The magnetic field induced by each of these cuboids is known analytically. Summing their contribution, we obtain a formula which only depends on the current amplitude in each of the wires. We have checked that this numerical model is in good agreement with the measured oscillation frequencies for various trap parameters.

On the experimental setup, the external bias field $\mathbf{B}_{\textrm{bias}}$ is obtained from four independent macroscopic pairs of coils surrounding the vacuum chamber (not shown in Fig.~\ref{fig:traps_and_axes}). For each pair, the current of the two coils flows in the same direction. Two of them are aligned along the $x$ axis and contribute to $B_{\textrm{bias},x}$ in an opposite way. This is necessary to either compensate partially the offset field created by the Z-wire along the $x$-axis, that would otherwise be too large and prevent reaching strong transverse trapping frequencies, or to increase the longitudinal field produced by the U-wires for reaching a value for $B_\textrm{min}$ large enough to avoid Majorana losses. The third pair of coils along the $y$-axis sets the value of $B_{\textrm{bias},y}$. The last pair is provided by the quadrupole coils themselves, which can be switched to a configuration where current in both coils flow in the same direction using electromechanical switches. In the latter configuration they are used to set $B_{\textrm{bias},z}$.

The experimental sequence used to calibrate the different components of the homogeneous magnetic field $\mathbf{B}_{\textrm{bias}}$ from the Zeeman shift of the magnetic sublevels is the following: we prepare a Bose-Einstein condensate and switch off abruptly all the currents contributing to the magnetic potential apart from the current producing the field component that we want to calibrate. After a time of flight of \SI{5}{\milli\second}, we switch on the microwave field for \SI{100}{\micro\second} and scan its frequency. The cloud free fall and expansion over this short time of flight is small with respect to the range over which the magnetic field is homogeneous. Since the atoms are trapped in the $|F=1,m_F=-1\rangle$ state, we can only couple them to the $|F=2,m_F=-2,-1,0\rangle$ states depending on the microwave polarization because of selection rules. The resonance frequencies are shifted due to the Zeeman effect $\Delta\nu=|g_F(m_1+m_2)| \mu_B B/h$ where $|g_F|=1/2$ is the hyperfine Land\'e g-factor, $\mu_B$ the Bohr magneton, $m_{1,2}$ the projections of the total spin for the initial and final states and $B$ is the modulus of the total magnetic field that remains after the trap switch off. Note that $B$ contains the contribution of the Earth magnetic field in addition to the field component that we want to calibrate. We finally record the population of the $F=2$ state with absorption imaging. We observe three resonances, from which we deduce $B$. Repeating the experiment for different currents in the external coils and fitting all the results together we deduce the relation between each magnetic-field component and the current in the external coils. Note that we also let the three components of Earth magnetic field (and possible other stray fields) as free parameters. The modulus of the magnetic field that we estimate for this contribution is 0.55\,G which is close to the Earth magnetic field (0.47\,G), such that stray fields at the position of the atoms remain limited. This method allows us to determine the magnetic-field components of $\mathbf{B}_{\textrm{bias}}$ with an uncertainty of 1\,mG.

\begin{figure}[t]
\centering
\includegraphics[width=.98\linewidth]{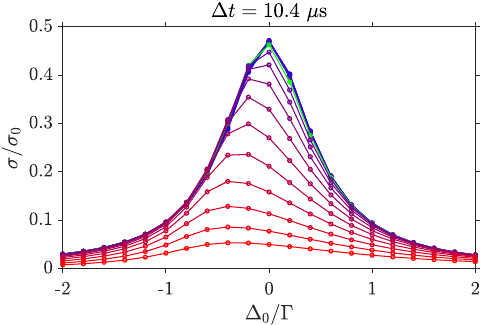}\\
\includegraphics[width=.98\linewidth]{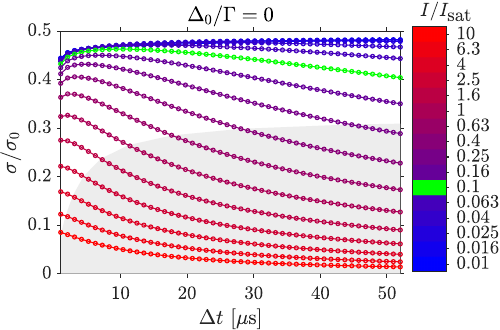}
\caption{\label{imaging} \textbf{Top} - Effective absorption cross section of the imaging beam in units of the resonance cross section $\sigma_0$ deduced from the optical Bloch equations for a fixed pulse duration as a function of the imaging beam detuning $\Delta_0$ in units of the linewidth $\Gamma$. \textbf{Bottom} - Same quantity for a fixed imaging beam detuning $\Delta_0/\Gamma=0$ as a function of the imaging pulse duration $\Delta t$. The shaded area indicates the region where the signal saturates the CCD sensor of the camera. In both graphs, each curve corresponds to a different imaging beam intensity.
}
\end{figure}

A similar technique is also used to measure $B_{\rm min}$ with a higher accuracy than the one provided by our modeling of the magnetic landscape. We shine the microwave field on the trapped atoms during \SI{200}{\micro\second} and wait for \SI{20}{\milli\second} before switching off the confinement in order to let the high-field seeker $F=2$ atoms be expelled by the magnetic potential. Scanning the frequency of the microwave field and detecting the remaining population in $F=1$, we can clearly distinguish three resonances corresponding to the different polarizations of the microwave field as shown in Fig.~\ref{three_pics}. The frequency difference between the resonances gives access to $B_{\rm min}$ with an uncertainty of 10\,mG.

\section{Simulations of the effective cross section for absorption imaging}
\label{calib_imaging}

Absorption imaging relies on the knowledge of the cross section $\sigma$ describing the interaction between the imaging beam and the atoms~\cite{Ketterle1999}. In the case of a two-level atom, we can write
\begin{equation}
    \sigma=\frac{\sigma_0}{1+4\left(\Delta/\Gamma\right)^2+I/I_{\textrm{sat}}}
    \label{eq:2-level_sigma}\textbf{}
\end{equation}
where $\sigma_0=\hbar\omega\Gamma/(2I_{\textrm{sat}})$ is the resonant cross section, $\omega/2\pi$ is the laser frequency, $\Delta=\omega-\omega_{\textrm{at}}$ is the laser detuning from the atomic resonance of frequency $\omega_{\textrm{at}}$, $\Gamma\simeq2\pi\times\SI{9.79}{\mega\hertz}$ is the natural linewidth of the excited state, $I$ is the laser intensity and $I_\textrm{sat}=\SI{6.26}{\milli\watt\per\centi\metre\squared}$ is the saturation intensity. 

Sodium atoms are alkali atoms. Their hyperfine structure encompasses 8 levels in the ground state. The first-excited state is a fine-structure doublet and only the $D_2$ line, including 16 levels, presents cycling transitions~\cite{Steck2010}. The excited-state structure is not very large compared with $\Gamma$ which makes the two-level atom approximation relatively inaccurate. Moreover, sodium atoms are quite light which means that the Doppler shift $\Delta_{\textrm{Dop}}\simeq2\pi\times\SI{50}{\kilo\hertz}$ associated with the absorption of a single photon is not negligible. This limits to a few hundreds the number of photons that can be scattered by an atom from the imaging beam before becoming off-resonant and in turn sets an upper bound on the pulse duration of a few \SI{}{\micro\second}.

To have a better insight of the performance of the absorption imaging setup, we have solved the optical Bloch equations taking into account the 24 electronic levels involved within the $D_2$ line. The Doppler shift due to the absorption of photons is taken into account through a time-dependent detuning $\Delta(t)$

\begin{equation}
    \Delta(t)=\Delta_0-\Delta_{\textrm{Dop}}\int_0^t\Gamma\rho_{ee}dt
\end{equation}
where $\Delta_0$ is the laser detuning with respect to the $F=2\rightarrow F'=3$ transition, and $\rho_{ee}=\sum_{F'=0\dots3}\sum_{m_{F'}=-F'}^{F'}\rho_{F',m_{F'};F',m_{F'}}$ corresponds to the total population of the excited states. Hence $\Gamma\rho_{ee}(t)$ is the total scattering rate.

In the experiment, the atoms occupy the $|F=1,m_F=-1\rangle$ state right before the magnetic trap is switched off. After time of flight, they are repumped to the $F=2$ states thanks to a short laser pulse at resonance with the $F=1\rightarrow F'=2$ transition. The repumper beam is oriented along the $x$ axis and circularly polarized. Right after this pulse, we shine another laser beam oriented along the $y$ axis and linearly polarized in the $(xz)$ plane. Relying on the model described above, we have calculated the effective cross-section for our imaging sequence. The results are shown in Fig.~\ref{imaging}. We see that the optimum imaging pulse duration depends on the laser intensity and that the largest effective cross-section is about 0.45$\sigma_0$. To maximize the signal to noise ratio without saturating the camera, we have to work with an intensity corresponding to a fraction of the saturation intensity and then the optimum pulse duration lies around \SI{10}{\micro\second}.

We have checked that the effective cross-section evaluated from these calculations is in good agreement with a careful calibration of the atom number performed from the expansion of Bose-Einstein condensates during a time of flight. The radii of the cloud scale according to equations derived in~\cite{castin:96} which are independant of the atom number. The initial radii however depends on the atom number through the chemical potential. Comparing the expansion of clouds with different atom number and originating from different trap geometry, we could check that the experimental effective cross-section is about 0.4$\sigma_0$.

\section{Scattering parameters of the coplanar waveguide}
\label{sec:cpw_S_pars}

\begin{figure}[t]
    \centering
    \includegraphics[width=\linewidth]{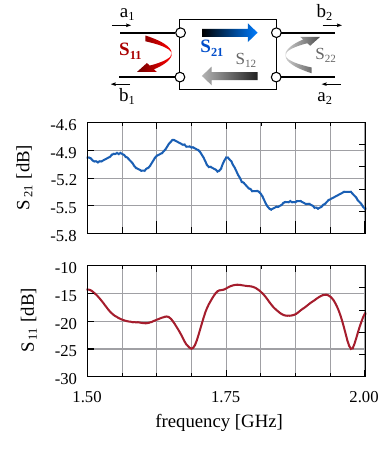}
    \caption{Upper frame: scattering parameters of a 2-port device. Middle and lower frames: measured $S_{21}$ and $S_{11}$ parameters of the on-chip coplanar waveguide (CPW).}
    \label{fig:cpw_Sij}
\end{figure}

Using a commercial vector network analyzer, we have characterized the spectral response of the full circuit of the coplanar waveguide by measuring its $S_{11}$ and $S_{21}$  scattering parameters (see Fig.~\ref{fig:cpw_Sij}). The amplitude reflection (transmission) parameter $S_{11}$ ($S_{21}$) corresponds to the ratio of the reflected (transmitted) signal to the incident signal when the output of the waveguide is perfectly matched: $S_{11} = \left. \frac{b_1}{a_1} \right| _{a_2 = 0}$ and $S_{21} = \left. \frac{b_2}{a_1} \right| _{a_2 = 0}$. The transmission of the coplanar waveguide is $-5.2~\pm~\SI{0.4}{\dB} $ within the range $1.5-\SI{2}{\giga\hertz}$. This modest value of power transmission ($\approx 30\%$) is attributed to losses in the in-vacuum leads, in the micrometric-sized gold wires that ensure connection to the chip's pads, and in the on-chip wires.

\begin{figure*}[t]
   \begin{minipage}[c]{.46\linewidth}
      \centering\includegraphics{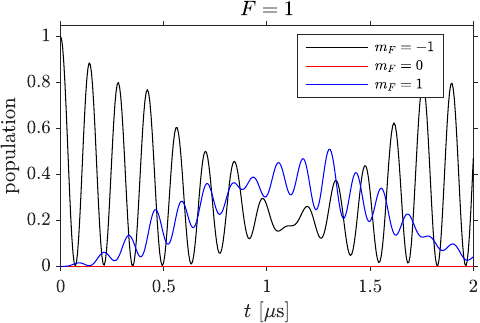}
   \end{minipage} \hfill
   \begin{minipage}[c]{.46\linewidth}
      \centering\includegraphics{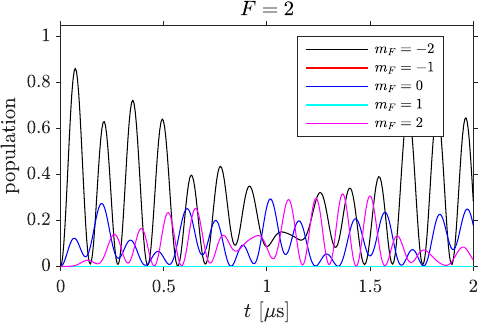}
   \end{minipage}
   \caption{\label{rwa}
\textbf{Left} - Coherent evolution of the population of the $\left | F=1,m_F\right\rangle$ states under the coupling with a microwave field taking into account the rotating wave approximation. We assume here that $B_\textrm{min}=1$\,G, $B_Y=5.4$\,G, $B_X=B_Z=0$\,G, and $\phi_X=\phi_Y=\phi_Z=0$ which corresponds to a pure linear polarization. The microwave frequency $\omega$ is chosen in order that the $\left | F=1,m_F=-1\right\rangle$ state is resonant with the $\left | F=2,m_F=-2\right\rangle$ state. All the atoms start in the $\left | F=1,m_F=-1\right\rangle$ state. \textbf{Right} - Same for the $\left | F=2,m_F\right\rangle$ states. 
}
\end{figure*}

\section{Microwave field coupling}
\label{sec:mwcoupling}

The Hamiltonian describing the coupling of a sodium atom in its ground state with a microwave field in the presence of a static magnetic field can be split into two terms
\begin{equation}
    \hat{H}(t)=\hat{H}_{\textrm{hfs}}+\hat{H}_{\textrm{B}}(t)
\end{equation}
where
\begin{align}
    \hat{H}_{\textrm{hfs}} &=\frac{\hbar\omega_\textrm{hfs}}{2}\frac{\hat{\mathbf{I}}\cdot \hat{\mathbf{J}}}{\hbar^2}\, ,\\
    \hat{H}_{\textrm{B}}(t) &= \frac{\mu_B}{\hbar}\left(g_J \hat{\textbf{J}}+g_I \hat{\textbf{I}}\right)\cdot\left(\textbf{B}_\textrm{s}+\textbf{B}_\textrm{mw}(t)\right).
\end{align}
Here, $\hat{\mathbf{I}}$ is the nuclear-spin operator, $\hat{\textbf{J}}$ the total electronic spin operator, $g_I$ the nuclear g-factor, $g_J$ the Landé g-factor, $\omega_\textrm{hfs}\simeq 2\pi\cdot \SI{1771.6}{\mega\hertz}$ the hyperfine splitting frequency, $\mu_B$ the Bohr magneton, $\textbf{B}_\textrm{s}$ the external static field and $\textbf{B}_\textrm{mw}(t)$ the microwave field. Since the contribution from $\hat{H}_{\textrm{B}}$ remains small compared with the hyperfine splitting energy in the situations we describe in this paper, we can write
\begin{align}
    \hat{H}_{\textrm{B}}(t) &\simeq \frac{\mu_B}{\hbar}g_F \hat{\textbf{F}}\cdot\left(\textbf{B}_\textrm{s}+\textbf{B}_\textrm{mw}(t)\right)
\end{align}
with $\hat{\textbf{F}}=\hat{\mathbf{I}}+\hat{\mathbf{J}}$ and $|g_F|=1/2$ the hyperfine Land\'e g-factor.

The most general expression for the microwave field can be written as
\begin{align}
    \mathbf{B}_\textrm{mw}(t) &= B_{X}\cos\left(\omega t+\phi_X\right)\mathbf{e}_X+B_{Y}\cos\left(\omega t+\phi_Y\right)\mathbf{e}_Y\notag\\
    &+B_{Z}\cos\left(\omega t+\phi_Z\right)\mathbf{e}_Z
\end{align}
where we have chosen the quantization axis $\mathbf{e}_Z$ along the static field $\textbf{B}_\textrm{s}=B_\textrm{min}\mathbf{e}_{x}$ such that $\mathbf{e}_X\equiv\mathbf{e}_z$, $\mathbf{e}_Y\equiv-\mathbf{e}_y$, $\mathbf{e}_Z\equiv\mathbf{e}_x$. We then introduce the unitary transform induced by the operator $\hat{U}_\textrm{hfs}(t)=\exp\left(-i \omega t \ \hat{H}_\textrm{hfs}/\hbar\omega_\textrm{hfs}\right)$. Applying $\hat{U}_\textrm{hfs}$ to $\hat{H}$ and neglecting all time-dependent terms according to the rotating wave approximation leads to
\begin{widetext}
\begin{align}
\hat{H}_{\textrm{eff}}=\hat{U}_\textrm{hfs}^\dagger \hat{H} \hat{U}_\textrm{hfs} -i\hbar \hat{U}_\textrm{hfs}^\dagger\frac{d \hat{U}_\textrm{hfs}}{dt} \simeq \frac{\hbar}{2} \left(\begin{smallmatrix}
-5\delta+2\delta_s & 0 & 0 & \sqrt{6}\Omega_-^* & \sqrt{3}\Omega_0^* & \Omega_+^* & 0 & 0 \\
0 & -5\delta & 0 & 0 & \sqrt{3}\Omega_-^* & 2\Omega_0^* & \sqrt{3}\Omega_+^* & 0 \\
0 & 0 & -5\delta-2\delta_s & 0 & 0 &  \Omega_-^* &  \sqrt{3}\Omega_0^* & \sqrt{6}\Omega_+^* \\
\sqrt{6}\Omega_- & 0 & 0 & 3\delta-4\delta_s & 0 & 0 & 0 & 0 \\
\sqrt{3}\Omega_0 & \sqrt{3}\Omega_-  & 0 & 0 & 3\delta-2\delta_s &0 & 0 & 0 \\
\Omega_+ & 2\Omega_0 & \Omega_- &  0 & 0 & 3\delta & 0 & 0\\
0 & \sqrt{3}\Omega_+  & \sqrt{3}\Omega_0 & 0 & 0 & 0 & 3\delta+2\delta_s & 0 \\
0 & 0 & \sqrt{6}\Omega_+ & 0 & 0 & 0 & 0 & 3\delta+4\delta_s
\end{smallmatrix}\right) \notag
\end{align}
\end{widetext}
where $\delta=(\omega_\textrm{hfs}-\omega)/4$, $\delta_s=|g_F|\mu_B B_\textrm{min}/\hbar$ and $\Omega_{0,+,-}=-|g_F|\mu_B B_\textrm{0,+,-}/\hbar$. We have introduced $B_+=(-B_{X}e^{-i\phi_X}+iB_{Y}e^{-i\phi_Y})/\sqrt{2}$, $B_-=(B_{X}e^{-i\phi_X}+iB_{Y}e^{-i\phi_Y})/\sqrt{2}$ and $B_0=B_{Z}e^{-i\phi_Z}$, the $\sigma_+$, $\sigma_-$, and $\pi$ components of the microwave field. The matrix is written in the basis $\left\{\left|F=1,m_F=-1\right\rangle\right.$, $\left|F=1,m_F=0\right\rangle$, $\left|F=1,m_F=1\right\rangle$, $\left|F=2,m_F=-2\right\rangle$, $\dots$, $\left. \left|F=2,m_F=2\right\rangle\right\}$. With these definitions, the microwave field can also be expressed as $\mathbf{B}_\textrm{mw}(t)=\dfrac{1}{2}\left[\mathbf{\mathcal{B}}e^{-i\omega t}+c.c.\right]$ with $\mathbf{\mathcal{B}}=B_+\mathbf{e}_++B_-\mathbf{e}_-+B_0\mathbf{e}_Z$ and $(\mathbf{e}_+,\mathbf{e}_-,\mathbf{e}_0)$ is the spherical basis: $\mathbf{e}_+=-(\mathbf{e}_X+i\mathbf{e}_Y)/\sqrt{2}$, $\mathbf{e}_-=(\mathbf{e}_X-i\mathbf{e}_Y)/\sqrt{2}$, $\mathbf{e}_0=\mathbf{e}_Z$.

In $\hat{H}_\textrm{eff}$, the off-diagonal coefficients directly correspond to the Rabi frequency $\Omega_R$ of the population oscillations for a given transition between two hyperfine states \cite{Bohi2010}. The amplitude $|\Omega_{0,+,-}|$ scales as $2\pi\times \SI{0.7}{\mega\hertz}\cdot$G$^{-1}$ with the modulus of the corresponding component of the microwave field $|B_{0,+,-}|$.

It is now straightforward to compute numerically the evolution of the density matrix over time and a typical result is shown in Fig.~\ref{rwa}, assuming a linear polarization of the microwave field, $\phi_X=\phi_Y=\phi_Z=0$, $B_Y=5.4$\,G and $B_X=B_Z=0$\,G. These parameters reproduces well the results observed in~Fig.~\ref{fig:rabi_high_power} when a Gaussian amplitude broadening is taken into account in the calculation. The calculation clearly shows that the decrease in the $\left|F=1,m_F=-1\right\rangle$ state at the maxima is due to an off-resonant coupling through the $\sigma^+$ line to the $\left|F=2,m_F=0\right\rangle$ state, which in turns populates the $\left|F=1,m_F=1\right\rangle$ state through a $\sigma^-$ line. Hence, at this large amplitude, the population dynamics cannot be reduced to the two states at resonance with the microwave field. Instead, the system is well described by a W configuration, with the five states $\left|F=2,m_F=-2\right\rangle$, $\left|F=1,m_F=-1\right\rangle$, $\left|F=2,m_F=0\right\rangle$, $\left|F=1,m_F=1\right\rangle$ and $\left|F=2,m_F=2\right\rangle$.

%\bibliography{biblio}


\end{document}